\documentclass[preprint2]{aastex}

\shortauthors{Klose et al.}
\shorttitle{Gamma-Ray Burst 000418}

\begin{document}

\title{The very red afterglow of GRB 000418 - further evidence for
       dust extinction in a GRB host galaxy\altaffilmark{1}}

\author{
S. Klose\altaffilmark{2},
B. Stecklum\altaffilmark{2},
N. Masetti\altaffilmark{3},
E. Pian\altaffilmark{3},
E. Palazzi\altaffilmark{3},
A. A. Henden\altaffilmark{4},
D. H. Hartmann\altaffilmark{5},
O. Fischer\altaffilmark{6},
J. Gorosabel\altaffilmark{7},
C. S\'anchez-Fern\'andez\altaffilmark{8},
D. Butler\altaffilmark{9},
Th. Ott\altaffilmark{9},
S. Hippler\altaffilmark{10}, 
M. Kasper\altaffilmark{10}, 
R. Weiss\altaffilmark{10},
A. Castro-Tirado\altaffilmark{8,11},
J. Greiner\altaffilmark{12},
C. Bartolini\altaffilmark{13}, 
A. Guarnieri\altaffilmark{13}, 
A. Piccioni\altaffilmark{13},
S. Benetti\altaffilmark{14}, 
F. Ghinassi\altaffilmark{14}, 
A. Magazz\'u\altaffilmark{14},
K. Hurley\altaffilmark{15},
T. Cline\altaffilmark{16},
J. Trombka\altaffilmark{16},
T. McClanahan\altaffilmark{16},
R. Starr\altaffilmark{16},
J. Goldsten\altaffilmark{17},
R. Gold\altaffilmark{17},
E. Mazets\altaffilmark{18}, 
S. Golenetskii\altaffilmark{18},
K. Noeske\altaffilmark{19}, 
P. Papaderos\altaffilmark{19},
P. M. Vreeswijk\altaffilmark{20},
N. Tanvir\altaffilmark{21},
A. Oscoz\altaffilmark{22}, 
J. M. Mun\~noz\altaffilmark{22},
J. M. Castro Ceron\altaffilmark{23}
}

\altaffiltext{1}{Based on observations collected at the Bologna
Astronomical Observatory in Loiano, Italy; at the TNG, Canary Islands, 
Spain; at the German-Spanish Astronomical Centre, Calar Alto, operated by 
the Max-Planck-Institut for Astronomy, Heidelberg, jointly with the 
Spanish National Commission for Astronomy; at the U. S. Naval 
Observatory and at the UK Infrared Telescope.}
\altaffiltext{2}{Th\"uringer Landessternwarte Tautenburg, D--07778 
                 Tautenburg, Germany.}
\altaffiltext{3}{Istituto Tecnologie e Studio Radiazioni Extraterrestri, CNR, 
                 Via Gobetti 101, I-40129 Bologna, Italy.}
\altaffiltext{4}{US Naval Observatory, Flagstaff station, P.O. Box 1149, 
                 Flagstaff, AZ 86002-1149.}
\altaffiltext{5}{Department of Physics and Astronomy, Clemson University, 
                 Clemson, SC 29634-1911.}
\altaffiltext{6}{Astrophysikalisches Institut und Universit\"ats-Sternwarte, 
                 Schillerg\"asschen 2, D--07745 Jena, Germany.}
\altaffiltext{7}{Danish Space Research Institute, Juliane Maries Vej 30,
                 DK--2100 Copenhagen, Denmark.}
\altaffiltext{8}{Laboratorio de Astrof\'{\i}sica Espacial y F\'{\i}sica 
                 Fundamental (LAEFF-INTA), P.O. Box 50727, E--28080 Madrid, 
                 Spain.}
\altaffiltext{9}{Max-Planck-Institut f\"ur Extraterrestrische Physik, 
                 Giessenbachstra\ss{}e, D--85748 Garching, Germany.}
\altaffiltext{10}{Max-Planck-Institut f\"ur Astronomie, K\"onigstuhl 17,
                  D--69117 Heidelberg, Germany.}
\altaffiltext{11}{Instituto de Astrof\'{\i}sica de Andaluc\'{\i}a (IAA-CSIC), 
                  P.O. Box 03004, E--18080 Granada, Spain.}
\altaffiltext{12}{Astrophysikalisches Institut Potsdam, D--14482 Potsdam, 
                  Germany.}
\altaffiltext{13}{Dipartimento di Astronomia, Universit\'a di Bologna, Via 
                  Ranzani 1, I-40127 Bologna, Italy.}
\altaffiltext{14}{Telescopio Nazionale Galileo, Centro Galileo Galilei, 
                  Calle Alvarez de Abreu 70, E-38700 Santa Cruz de La Palma, 
                  Canary Islands, Spain.}
\altaffiltext{15}{University of California at Berkeley, Space Sciences 
                  Laboratory, Berkeley, CA 94720-7450.}
\altaffiltext{16}{NASA Goddard Space Flight Center, Greenbelt, MD 20771.} 
\altaffiltext{17}{Applied Physics Laboratory, Johns Hopkins University, 
                  Laurel, MD 20723.}
\altaffiltext{18}{Ioffe Physico-Technical Institute, St. Petersburg, 194021 
                  Russia.}
\altaffiltext{19}{Universit\"ats-Sternwarte, Geismarlandstrasse 11, 
                  D-37083 G\"ottingen, Germany.}
\altaffiltext{20}{Astronomical Institute 'Anton Pannekoehk', 
                  Kruislaan 43, 1098 SJ Amsterdam, The Netherlands.}
\altaffiltext{21}{Joint Astronomy Centre, 660 N. A'ohoku Place, Hilo, 
                  Hawaii 96720.}
\altaffiltext{22}{Instituto de Astrof\'{\i}sica de Canarias, La Laguna, 
                  Tenerife, Spain.}
\altaffiltext{23}{Real Observatorio de la Armada, San Fernando Naval, Cadiz, 
                  Spain.}


\begin{abstract}

We report near-infrared and optical follow-up observations of the
afterglow of the Gamma-Ray Burst 000418 starting 2.5 days after the
occurrence of the burst and extending over nearly seven weeks. GRB 000418
represents the second case for which the afterglow  was initially
identified by  observations in the near-infrared. During the first 10
days its $R$-band afterglow was well characterized by a single
power-law decay with a slope of 0.86.  However, at later times the
temporal evolution of the afterglow flattens with respect to a simple
power-law decay. Attributing this to an underlying host galaxy we
find its magnitude to be $R$=23.9 and an intrinsic afterglow decay slope
of 1.22. The afterglow was very red with $R-K\approx$4 mag. 
The observations can be explained by an adiabatic, spherical fireball 
solution and a heavy reddening due to dust extinction in the host galaxy.
This supports the picture that (long) bursts are associated with 
events in star-forming regions.

\end{abstract}

\keywords{gamma-rays: bursts}


\section{Introduction}

The detection of Gamma-Ray Burst (GRB) afterglows still constitutes a
challenge to observational astronomy. This is illustrated by the fact
that three years after the discovery of GRB afterglows in the optical
(van Paradijs {\it et al.} 1997) less than 20 such optical
transients have been detected and followed up (see Greiner  2000 for
continuous updates). Among the afterglows recorded to date the
emission from GRB 000418 represents the second case for which the
afterglow was initially identified in a gamma-ray burst error box by
observations in the near-infrared (the first case is GRB 990705;
Masetti {\it et al.} 2000).

GRB 000418 was observed on April 18.41 UT by the third interplanetary
network (IPN), with Ulysses (Hurley {\it et al.} 1992) and the
Near-Earth Asteroid Rendezvous mission (NEAR, XGRS detector, Goldsten
{\it et al.} 1997) as its distant points, and GGS-Wind Konus (Aptekar
{\it et al.} 1995) close to Earth. The burst had
a duration of $\sim$30 seconds, and a fluence in the 25--100 keV range
of $\sim1.3\,\times\,10^{-5}$ erg cm$^{-2}$ (Hurley, Cline, \& Mazets
2000). These properties characterize the event as a long, relatively
bright GRB (cf. Fishman \& Meegan 1995).

Rapid follow-up and second epoch near-infrared (NIR) observations of
the central part of the 35 arcmin$^2$ sized GRB error box led to the
detection of a NIR source (Klose {\it et al.} 2000) at R.A.(J2000) =
12$^h$25$^m$19\fs30, decl.(J2000) = 20$^\circ$06$'$11\farcs6
($\pm$0\farcs5),  whose fading nature was quickly confirmed by
follow-up observations in the optical (Mirabal {\it et al.}  2000b).
The fading of the afterglow is clearly illustrated in Fig. 1 which
reports $R$-band images taken at TNG on April 20 and June 2, 2000.
The successful detection of the afterglow of GRB 000418 makes this an
afterglow of an ``IPN-only'' burst, with no additional available X-ray
(BeppoSAX) or \it BATSE \rm spectral information (GRB 991208 was the
first of these bursts with detected optical afterglow;
Hurley {\it et al.} 2000; Castro-Tirado {\it et al.} 2000). Triggered
by the NIR detection of  the afterglow of GRB 000418, many
observatories world-wide directed their attention to this burst. Here
we report on a subset of these activities, with emphasis on the
$R$-band light curve.

\section{Observations and data reduction}

The IPN error box of GRB 000418 became known two days after the burst.
The observations leading to the discovery of the GRB afterglow  were
performed on 2000 April 20 and April 22 with the 3.5-m telescope on
Calar Alto, Spain, equipped with the near-infrared camera \it Omega
Cass \rm (Lenzen {\it et al.} 1998) which was operated in the
polarimetric mode. The central part of the 
elongated GRB error box was imaged with the 3.5-m telescope in the
course of a Target of Opportunity project.  In the polarimetric
wide-field mode the \it Omega Cass \rm camera has a $5'\,\times\,5'$
field of view and an image scale of 0.3$''$/pixel. The  limiting
magnitude of the $K'$-band image is $\sim$18.5 for unpolarized objects
after adding all frames taken at different position angles of the
wire-grid polarizer.

Further near-infrared observations were performed at the Calar Alto
1.23-m telescope using the NIR camera \it MAGIC\rm, with the United
Kingdom Infrared Telescope (UKIRT) equipped with the UKIRT Fast-Track
Imager and using the Calar Alto 3.5-m telescope equipped with the \it
Omega Prime \rm near-infrared camera (Table 1). Optical follow-up observations
were performed at the Italian {\it Telescopio Nazionale Galileo}
(TNG),  La Palma, Spain, at the U.S. Naval Observatory Flagstaff
Station (NOFS), at the 1.52-m ``G.D. Cassini" telescope of the
University of Bologna in  Loiano, Italy, and on Calar Alto using the
3.5-m telescope with the multi-purpose instrument \it MOSCA\rm. TNG is
equipped with the OIG camera, which contains two EEV CCDs. The Loiano
telescope employs the BFOSC instrument, equipped with an EEV CCD. The
NOFS observations were performed with either the 1.3-m automated
telescope, equipped with a SITe CCD, or the 1.0-m telescope, equipped
with a Tektronix CCD.

The $K'$-band measurements performed on April 20.90 UT were reduced in
a standard fashion. The sky contribution was obtained from the median
of the dithered images and, in combination with a proper dark frame,
used to derive the flat field.  The images were sky-subtracted,
flat-fielded, bad-pixel corrected, and finally mosaicked. This
procedure was performed for each orientation of the polarimetric
analyzer. The final image is the sum of the frames obtained for each
orientation of the polarizer. Unfortunately, it was not possible to
derive the linear polarization from these frames.  The Calar Alto
3.5-m data of April 22 were obtained during variable sky and required
another method to estimate the flat field. The considerable temporal
variation of the sky background allowed the derivation of both the
flat field and bias by pixelwise linear regression, similar to the
scheme proposed by Fixsen, Moseley, \& Arendt (2000). Otherwise,  the
image processing was the same.

We determined the $K'$-band magnitude of the afterglow by means of 
optical photometry of secondary standard stars in the field around the
burster (Henden 2000) and by means of the USNO-SA2.0 Catalog (Monet
{\it et al.} 1996). The basic assumption in our procedure is that the
selected field stars are main-sequence stars and that their optical
colors are representative for their spectral type.  We corrected our
$K'$-band data for differences in the $K,K'$ magnitudes using the
relation $K'-K=0.20 (H-K)$ (Wainscoat \& Cowie 1992). The $H-K$ color
of the afterglow was calculated based on the $R-K'$ color measured on
April 20.9 UT (Table 1) and assuming a $F_{\nu} \propto
\nu^{-\beta}$ optical-NIR spectrum (see below); this gives 
$K'-K\approx$0.2 mag. Finally,
we used the wide-field $K'$-band image obtained at Calar Alto on May
15 to calibrate the narrow-field UKIRT $K$-band image from May
4. Although the photometry  of our NIR data is not as accurate as our
$R$-band photometry this  does not affect the basic content of the
following discussion (\S3). The $R$-band light curve is shown in 
Fig. 2.

Because GRB 000418 was at high Galactic latitude, we assume that the
faint photometric  reference stars exhibit the same Galactic
extinction as the GRB afterglow. The Schlegel, Finkbeiner, \& Davis
(1998) extinction maps predict $E(B-V)=0.04$ through the Galaxy along
Galactic coordinates $(l,b)$ = 265\fdg1, 80\fdg5. Assuming a ratio of
visual to selective extinction of 3.1, this gives $A_V\sim$0.12
mag. The dust distribution model of Chi \& Wolfendale (1991)  predicts
a larger visual extinction of $A_V\sim$0.19 mag, while the compilation
of Hakkila {\it et al.}  (1997) predicts a very small value of
$A_V\sim$0.02  mag. We take  the value derived from Schlegel {\it et
al.}, which is close to the mean of the latter two estimates. Then,
according to Rieke \& Lebofsky (1985), this corresponds to $A_R=0.09$
mag and $A_K$=0.01 mag.

For the $R$-band optical frames, psf-fitting was performed using
DAOPHOT-II as implemented in IRAF.  The deep frames provided a means
to find isolated, well-sampled stars in a field highly contaminated by
galaxies.  These stars were used to calculate mean point spread
functions, and the instrumental magnitudes of the optical transient
and several comparison stars were extracted.  The comparison star
standard $R$-band magnitudes were obtained from the field photometry
file of Henden (2000), and the afterglow magnitude determined
differentially with respect to the ensemble mean.

\section{Results and Discussion}

GRB 000418 is a ``IPN-only'' burst, and its afterglow was not observed
in the X-ray band. Moreover, radio observations of GRB 000418 were not
performed before April 29 (Frail 2000). The lack of these data limits
the diagnostic power of  afterglow observations in this case, but the
optical decay still yields a significant parameter (the slope of the
decay law), and future changes in slope might yet provide evidence for
beaming or contamination by light from an underlying supernova.
Therefore, we concentrate here on the interpretation of the
optical/NIR afterglow of this burst. We follow the ``standard
procedure'' (e.g., Wijers, Rees, \& M\'esz\'aros 1997; Galama {\it et
al.} 1998; Frail, Waxman, \& Kulkarni 2000)  for interpreting
afterglows within the context of the so-called fireball model. In this
widely accepted model for GRB afterglows, the spectral signature is
due predominantly to synchrotron emission from relativistic electrons
accelerated in the strong shock wave that propagates into the burst
environment (and also a reverse shock propagating into the burst
ejecta, in close analogy to the situation  in supernova remnants, with
the exception that GRBs produce highly relativistic shocks). The
emerging spectrum is a combination of emission from internal shocks
(energy dissipation within the ejecta) and external shocks driven into
a medium that could be roughly uniform (ISM-like) or have strong
radial profiles, as would be the case if the external medium is wind
material from a massive star prior to becoming a collapsar. The basic
properties of these relativistic shock models are reviewed by Piran
(1999) and M\'esz\'aros (2000). In the simplest model (an impulsive
blast wave, a uniform circumburster medium, and spherical symmetry),
the evolution of the afterglow flux density evolves according to a
power-law in time, i.e., $F_\nu (t) \sim t^{-\alpha}\nu^{-\beta}$ once
the synchrotron peak has moved through the  bandpass. The
time-dependent apparent magnitude in a given photometric band is then
described by $m(t_2) - m(t_1) = 2.5 \alpha$\, $\lg(t_1/t_2)$, where
$t$ is measured from the time of the occurrence of the burst (see Vrba
{\it et al.} 2000 for a discussion of the zero point of $t$). Possible
deviations from a power-law decay of the afterglow flux can be caused
by a variety of reasons (see Rhoads \& Fruchter 2000 for a recent
discussion): different afterglow physics and model assumptions about
the environment (cf. Panaitescu, M\'esz\'aros, \& Rees 1998), a jet
geometry (cf. Sari, Piran, \& Halpern 1999),  supernova light which
adds to the afterglow flux (cf. Bloom {\it et al.} 1999), light echoes
due to dust clouds around the burst (e.g., Esin \& Blandford 2000) or
an emerging host galaxy which becomes  detectable when the afterglow
is faint enough (cf. Zharikov, Sokolov, \& Baryshev 1998). Most
afterglows are very faint in the optical, even at early times, and
therefore are usually observed at the detection limit of optical
telescopes. An observational search for deviations from the prediction
of the simplest fireball model is therefore difficult in most
cases. GRB 000418 is no exception to this.

If we consider only the first five data points of the afterglow light
curve (Fig.~2), then the best-fit temporal power-law slope is
$\alpha=0.86\pm0.06$ ($\chi^2$/dof=0.31), in agreement with Mirabal
{\it et al.}  (2000a,b). However, starting in May (after 10 days
since the GRB trigger) the
$R$-band data indicate a significant flattening with respect to the predicted
power-law decay of the afterglow flux with $\alpha$=0.86. 
Light from an underlying host galaxy can explain this discrepancy.
The best fit now is $\alpha$=1.22 and $R_{\rm host}=23.9$
($\chi^2$/dof=2.18). (In this fit we 
have omitted the single Calar Alto $R$-band data
point from June 2. If we include it we get $\alpha$=1.29, 
$R_{\rm host}=23.8$.)

At the time of the $K'$-band observation (April 20.9 UT) the afterglow
was very red with $R-K = (3.9\pm0.2)$ mag (Table 1).  A similar red
color was observed for GRB 980329 (Palazzi  {\it et al.} 1998) and GRB
990705 (Masetti {\it et al.} 2000). If we take reddening by Galactic
dust into account (\S2), then, on April 20.9 UT, the observed spectral
slope across the $R-K$-band is $\beta=1.90\pm0.15$, where the
uncertainty is dominated by the error of the $K'$-band
magnitude. The slow power-law decay of the afterglow, $\alpha=1.22$,
favors spherically symmetric evolution (Table 2). Following the
analysis of Sari, Piran, \& Narayan (1998) for a shock expanding into
a uniform medium, the predicted value for the spectral slope $\beta$
across the $R-K$ passband is $(2\alpha-1)/3=0.48$ or $2\alpha/3=
0.81$, respectively, depending on whether or not the bulk of the
radiating electrons is in the fast cooling regime, i.e. whether or not
the  characteristic cooling time is shorter than the age of the shock.
Since our observations started several hours after the GRB we expect
that at this time the bulk of the electrons was already 
in the slow cooling regime. The slope $p$ of the 
power-law distribution of electron Lorentz factors,  $N(\gamma)
d\gamma \sim \gamma^{-p}$, is then $p=4\alpha/3+1=2.63$. The fact that
the observed spectral index $\beta$ significantly exceeds the predicted value
could plausibly be a consequence of extinction in the host galaxy.

Extinction in the host galaxy may change the spectral energy
distribution (SED) of the afterglow in the optical/NIR bands such that
the SED in the optical/NIR range can no longer be approximated by a
single power-law. This renders the color estimate redshift-dependent.
Assuming  that the local extinction in the GRB environment is
dominated by dust, we corrected the observed optical/NIR spectral slope
of the afterglow by adopting a local optical extinction curve
$A(\lambda$) in the GRB environment which is
similar to that for the Galactic interstellar dust
(S\={u}d\v{z}ius, Bobinas, \& Raudeli\={u}nas 1996). However, we let
the  ratio of total-to-selective extinction, $R_{\rm dust}$, a free
parameter, in order to describe approximately the various extinction
curves found in star-forming regions in our Galaxy (e.g., Vrba, Coyne, 
\& Tapia 1993) and in starburst galaxies (e.g., Calzetti 1997).  For a redshift
of $z=1.118$ (Bloom {\it et al.} 2000)  and a Galactic visual
extinction in the direction of the optical transient of $A_V$ = 0.12
mag (\S2) we find that a visual extinction in the host galaxy of
$A_V^{\rm host} \sim (0.96\pm0.2)\, R_{\rm dust}$/3.17 mag is
required in order to change the spectral index from  $\beta=0.81$ to
the observed one, assuming no contribution from the underlying host
galaxy. We thus conclude that the observed optical/NIR color can in
principle be forced  to agree with the observed power-law decline
slope {\it if}, on April 20.9 UT, the observations were carried out at
a time when the effects of jet formation were not manifest, i.e. the
light curve is well represented by the adiabatic, spherical fireball
solution  (e.g., Wijers {\it et al.} 1997; Sari {\it et al.} 1998),
and the afterglow was significantly reddened due to dust extinction in
the host galaxy. While this argument supports the notion that the GRB
was associated with a massive star  embedded in a star-forming region
(e.g., Paczy\'nski 1998; Fryer, Woosley, \& Hartmann 1999),  the
optical/NIR observations alone cannot constrain the position of the
intervening dust in the GRB host galaxy. We note, however,  that our
value for $A_V^{\rm host}$ is comparable to those deduced from
observations of the optical afterglows of GRB 980703  (Castro-Tirado
{\it et al.} 1999b; Vreeswijk {\it et al.} 1999a), GRB 971214 (Dal 
Fiume {\it et al.} 2000; Halpern
{\it et al.} 1998; Ramaprakash  {\it et al.} 1998), and GRB 980329
(Palazzi {\it et al.} 1998).

Combining the May 3.3 UT $R$-band measurement with the May 4.4 UT
$K$-band  data gives a color of the afterglow of $R-K=3.0\pm$0.4 mag.
Within our measurement errors this is not in conflict with our April
20.9 data, assuming that the color is now affected by the color
of the underlying host galaxy. On May 8.9
UT (day 20.5) we measure $V-R$=0.64 which also indicates that  the
optical spectrum is less steep at later times. The $R-K$ color
value for the afterglow, together with the
assumption of a flat spectrum for the host, suggests that on May 8.9 the
main contribution to the measured $V$ magnitude is due to the host galaxy,
as we derive $V_{\rm host} \sim 24.2$ (and $K_{\rm host} \sim 22.1$)
in this hypothesis. Our imaging, however, does not reveal the GRB
host galaxy as an extended structure. The source remaines point-like.
We have compared the TNG image from April 20 with the TNG image
obtained on May 23 in order to get an information about the
displacement $d$ of the GRB afterglow from the luminosity  center of
the host galaxy. In the TNG image from April 20 the 
point spread function (psf) of the source 
is mainly determined by the GRB afterglow, whereas in the
TNG image from May 23 it is mainly determined by the host galaxy.
This allowed us to check if there is evidence for a shift of the
center of the psf with respect to a near-by
reference star. We found no evidence for such a shift and conclude 
$d<$0\farcs1.

Initially, the  observed temporal power-law decay of the afterglow of
GRB 000418 was somewhat  similar to that observed in the early light
curve of GRB 990510 (Table 2), the  time scale however was notably
different. The afterglow of GRB 990510 was slow for only about the
first 24 hours, afterwards steepening to $\alpha >2$. By contrast, the
afterglow of GRB 000418 displayed this slow decay mode for  the first
10 days  (see Vrba {\it et al.} 2000 for a compilation of other
late-time temporal power-law slopes). With the underlying host galaxy
becoming apparent in the afterglow flux the deduced temporal slope
$\alpha=1.22$ brings this burst now close to GRBs 970228, 970508
(Garcia {\it et al.} 1998), and 971214 
(Kulkarni {\it et al.} 1998; Ramaprakash  {\it et al.} 1998).

\section{Conclusions}

Based on the $R$-band data alone the parameters characterizing a
GRB afterglow are not strongly constrained. 
In general, the two observational
parameters easiest to determine in the optical with reasonable accuracy are the
temporal power-law decay slope, $\alpha$, and the redshift of the GRB
host, if present. The spectral slope, $\beta$, is affected by
extinction and thus much less  accurately known, and model-dependent.
On the other hand, even the deduced parameter $\alpha$ can be
misleading if the observed temporal evolution of the afterglow flux is
affected by an underlying host galaxy which becomes visible only at
later times (cf. Castro-Tirado \& Gorosabel 1999). The database
available to us is not sufficient to distinguish between the various
non-standard afterglow models, like a blast into a uniform medium
vs. a shock expanding into a pre-existing stellar wind with a
circumburster density profile $\rho \propto r^{-2}$, or the possible
effects from anisotropic outflows (M\'esz\'aros, \& Rees, \& Wijers
1998; Chevalier \& Li 1999).  In particular, the  GRB afterglow
decline slope $\alpha$ is not very steep, which hampers an unambiguous
detection of a supernova component in the afterglow (cf. Bloom {\it et
al.} 1999).  The observational evidence which we found for a reddening
of the  afterglow in the GRB host galaxy supports however the picture
that the GRB was associated with a dusty star forming region.

The redshift of GRB 000418 has been estimated from a line doublet
attributed to [OII] emitted by the host galaxy (Bloom {\it et al.}
2000). With this preliminary identification the redshift is $z$=
1.118, which is close to the mean of the current redshift database.
Assuming spherical emission, the energy release in gamma-rays of GRB
000418 was thus about $5\,\times\,10^{52}$ erg (Bloom {\it et al.}
2000) and the total observed fluence in the $R$-band between day 2.5
and day 50 is about 0.03\% of the energy  in the gamma-ray regime. The
energy release in gamma-rays is about two orders of magnitude less
luminous than the current record for  GRB 990123  (Andersen {\it et
al.} 1999; Kulkarni {\it et al.} 1999).  It   places this burst near
the median of the dozen presently determined energy releases (see
Klose 2000 for a recent review). If the brightest bursts are those
with the smallest collimation angles (in order to reduce the energy
requirements) GRB 000418 might not be strongly beamed. Indeed, based
on our data, covering day 2 to day 50 after the burst, we do not find
evidence for a noticeable break in the light curve, which would be the
telltale signature of an emerging view of the jet's edge.  Moreover,
within the context of the fireball model  (e.g., Wijers {\it et al.}
1997; Sari {\it et al.} 1998) we can explain the optical/NIR data with
the simple spherical solution. Since the observations of this
afterglow  do not cover the first two days, we are unable to state
with certainty  that we have not missed any break in the early light
curve. However, if  that had been the case the early power-law decay
would have been even slower, i.e. $\alpha<1$, which would make it
close to that of GRB 970508 (e.g., Castro-Tirado {\it et al.}  1998)
and perhaps place the spectrum  in the $F_\nu \sim \nu^{1/3}$ regime
with $-1/2 < \alpha < 1/3$ (Sari {\it et al.} 1998). 

GRB 000418 is only the $\sim$10th burst for which a well observed
afterglow lightcurve has been established. This case continues to
support the  basic concepts of the fireball scenario, but it also
reminds us that every afterglow is unique and that a large database
will be required  to establish the dynamic range of afterglow
properties.

\vspace*{5mm}

Note added in manuscript: When this paper was submitted Bloom {\it et
al.} (2000, GCN Circ. 689) and  Metzger {\it et al.} (2000, GCN
Circ. 733) reported the detection of the GRB host galaxy at $R =
24\pm0.3$ and $R = 23.9\pm0.2$, respectively, in  agreement with our
fit to the lightcurve.



\acknowledgements SK acknowledges valuable comments by Fred Vrba,
Flagstaff, and  Attila M\'esz\'aros, Prague.  J. Gorosabel
acknowledges  the receipt of a Marie Curie Research Grant from the
European Commission. This work has been  partially supported by the
Spanish INTA grant Rafael Calvo Rod\'es and  the Spanish CICYT grant
ESP95-0389-C02-02.  KH is grateful for Ulysses support under JPL
Contract 958056, and for NEAR support under NAG 5 9503.  We thank the
staff of the  Calar Alto observatory for performing service
observations. We thank the referee, D. Helfand, for his constructive
remarks.


\newpage

\onecolumn
 

\begin{figure}
\vspace*{2mm}
\epsscale{1}
\plotone{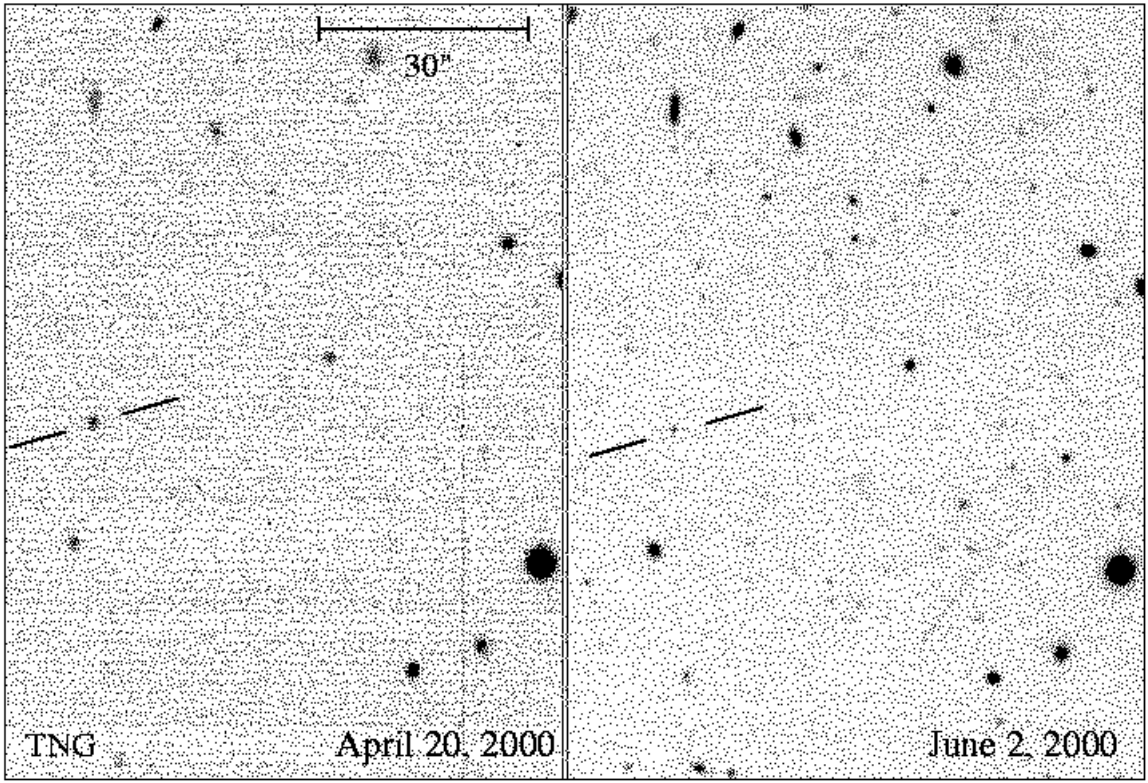}
\vspace*{0.5cm}
\caption{
Finding chart for the field of GRB 000418. The afterglow is
indicated. Left: $R$-band image taken with the TNG telescope on April
20.9 UT. Right: $R$-band image taken with TNG  on June 2.9 UT (Table
1). North is up, East is left.}
\end{figure}


\begin{figure}
\vspace*{-4cm}
\epsscale{1.1}
\plotone{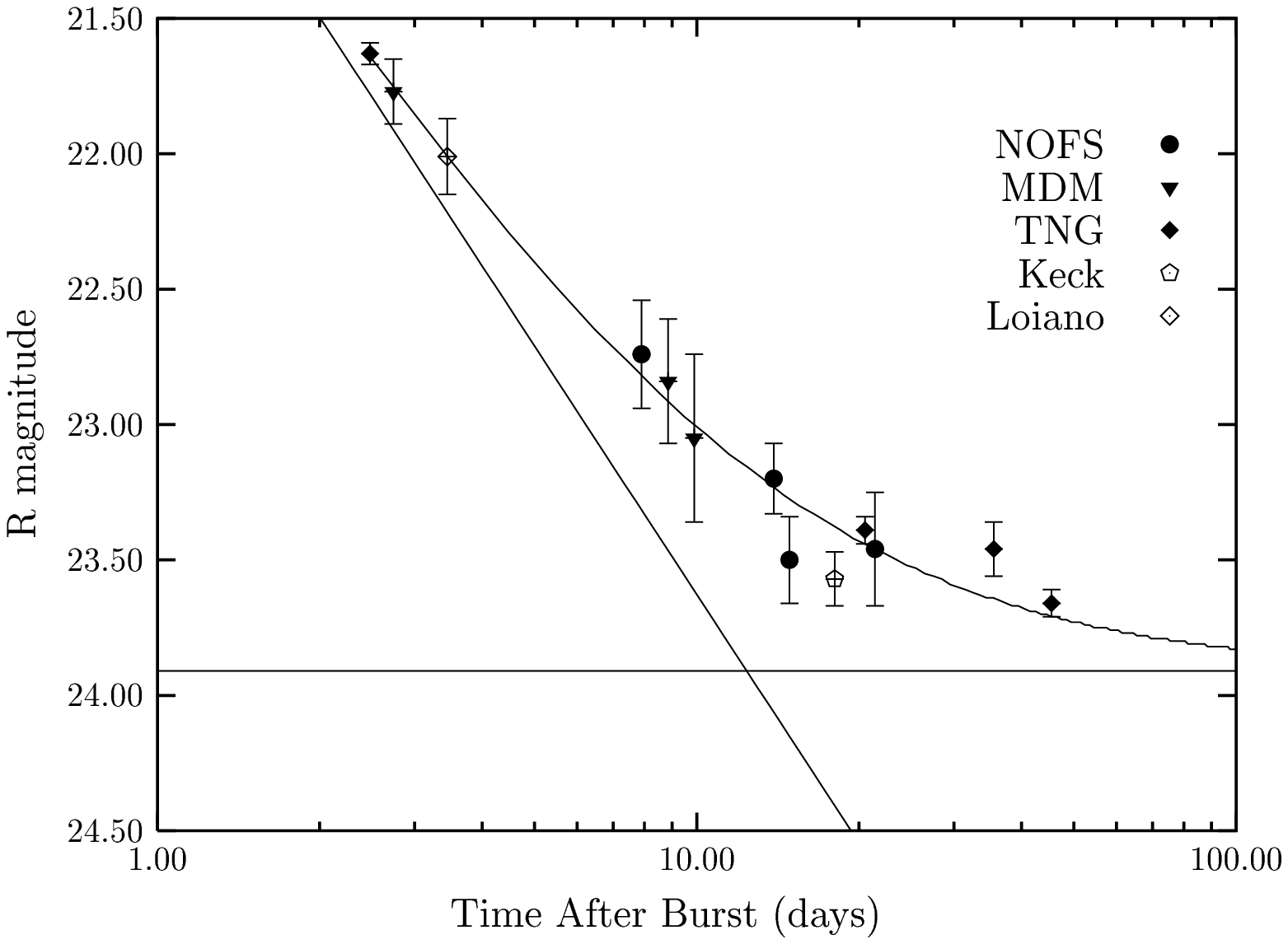}
\vspace*{-10cm}
\caption{
The $R$-band light curve of GRB 000418. The solid line is the
best fit to the data which predicts  an underlying host galaxy with
$R$=23.9. The straight lines show the individual contributions of the
GRB afterglow  (following a power-law decay with $\alpha=1.22$) and
the host galaxy to the total observed flux.}
\end{figure}

\newpage

\begin{deluxetable}{rrrlllr}
\tablewidth{15cm}
\tablecaption{Summary of observations of GRB 000418\tablenotemark{a}}
\tablehead{
\colhead{Date}      &
\colhead{$t-t_0$\tablenotemark{b}} &
\colhead{Telescope\tablenotemark{c}} & 
\colhead{Eposure (s)}    & 
\colhead{Filter}    & 
\colhead{Magnitude} &
\colhead{Ref.}}
\startdata
 Apr 20.89  UT & 2.48 &TNG 3.5-m          &  600             & $R$  &  21.63$\pm$0.04   & 1 \\ 
 Apr 20.90  UT & 2.49 &Calar Alto 3.5-m   & 1200             & $K'$ &  17.5$\pm$0.50\tablenotemark{d,e}& 2 \\        
 Apr 20.93  UT & 2.52 &Calar Alto 1.23-m  & 3600             & $K'$ &  17.9$\pm$0.20    & 1 \\      
 Apr 20.94  UT & 2.53 &Loiano 1.52-m      & 3$\,\times\,$1200& $R$  &$>20.5$            & 1 \\      
 Apr 21.15  UT & 2.74 &MDM 2.4-m          &                  & $R$  &  21.77$\pm$0.12\tablenotemark{f} & 3 \\     
 Apr 21.86  UT & 3.45 &Loiano 1.52-m      & 3$\,\times\,$1200& $R$  &  22.01$\pm$0.14   & 1 \\ 
 Apr 21.94  UT & 3.53 &Calar Alto 1.23-m  & 1350             & $J$  &  $>$18            & 1 \\          
 Apr 22.12  UT & 3.71 &Calar Alto 3.5-m   & 1200             & $K'$ &  $>$18.3\tablenotemark{e}        & 1 \\         
 Apr 24.92  UT & 6.51 &Calar Alto 1.23-m  & 4095             & $K'$ &  $>$17.5          & 1 \\         
 Apr 26.32  UT & 7.91 &NOFS 1.3-m         & 3$\,\times\,$600 & $R$  &  22.74$\pm$0.20\tablenotemark{f} & 4 \\       
 Apr 27.26  UT & 8.85 &MDM  2.4-m         &                  & $R$  &  22.84$\pm$0.23\tablenotemark{f} & 5 \\       
 Apr 28.3   UT & 9.89 &MDM  2.4-m         &                  & $R$  &  23.05$\pm$0.31\tablenotemark{f} & 6 \\       
 May 02.31  UT &13.90 &NOFS 1.3-m         & 21$\,\times\,$600& $R$  &  23.20$\pm$0.13   & 1 \\           
 May 03.26  UT &14.85 &NOFS 1.3-m         & 36$\,\times\,$600& $R$  &  23.50$\pm$0.16   & 1 \\           
 May 04.44  UT &16.03 &UKIRT 3.8-m        & 3240             & $K$  &  20.50$\pm$0.4    & 1 \\   
 May 06.42  UT &18.01 &Keck I 10-m        &                  & $R$  &  23.57$\pm$0.10   & 7 \\    
 May 08.89  UT &20.48 &TNG 3.5-m          & 3$\,\times\,$600 & $R$  &  23.39$\pm$0.05   & 1 \\     
 May 08.92  UT &20.51 &TNG 3.5-m          & 3$\,\times\,$600 & $V$  &  24.03$\pm$0.07   & 1 \\     
 May 09.82  UT &21.41 &NOFS 1.0-m         &12$\,\times\,$1200& $R$  &  23.46$\pm$0.21   & 1 \\ 
 May 11.95  UT &23.54 &Calar Alto 3.5-m   & 2100             & $J$  &  $>$19.5          & 1 \\ 
 May 15.01  UT &26.60 &Calar Alto 3.5-m   & 840              & $J$  &  $>$20.5          & 1 \\ 
 May 15.02  UT &26.61 &Calar Alto 3.5-m   & 1680             & $K'$ &  $>$19.5          & 1 \\   
 May 23.93  UT &35.52 &TNG 3.5-m          & 3$\,\times\,$900 & $R$  &  23.46$\pm$0.10   & 1 \\  
 Jun 02.88  UT &45.47 &Calar Alto 3.5-m   & 4$\,\times\,$600 & $R$  &  23.41$\pm$0.08   & 1 \\  
 Jun 02.91  UT &45.50 &TNG 3.5-m          & 3$\,\times\,$1200& $R$  &  23.66$\pm$0.05   & 1 \\  
\enddata
\tablenotetext{a}{supplemented by data taken from the literature}
\tablenotetext{b}{$t_0$ is the time of the occurrence of the burst:
 April 18.41 UT}
\tablenotetext{c}{TNG = \it Telescopio Nazionale Galileo, \rm La Palma, Spain;
                 NOFS = U.S. Naval Observatory Flagstaff Station;
               Loiano = \it G.D. Cassini \rm telescope of the University of Bologna,
                        Loiano, Italy;
                  MDM = Hiltner telescope at the MDM Observatory on Kitt Peak;
                UKIRT = United Kingdom Infrared Telescope, Mauna Kea, Hawaii}
\tablenotetext{d}{affected by a bad focus}
\tablenotetext{e}{reduced sensitivity due to polarimetric mode}
\tablenotetext{f}{corrected according to improved photometry of 
 secondary standard stars by Henden (2000)}
\tablerefs{
(1) this paper;
(2) Stecklum {\it et al.} 2000;
(3) Mirabal {\it et al.} 2000;
(4) Henden {\it et al.} 2000; 
(5) Mirabal, Halpern, \& Wagner 2000a; 
(6) Mirabal, Halpern, \& Wagner 2000b;
(7) Metzger \& Fruchter 2000}
\end{deluxetable}


\newpage 

\begin{deluxetable}{clrlllr}
\tablewidth{15cm}
\tablecaption{$R$-band GRB afterglows with well observed power-law breaks}
\tablehead{
\colhead{GRB} &
\colhead{redshift}   & 
\colhead{Ref.}   & 
\colhead{$\alpha_1$}   & 
\colhead{$t_{\rm break}[{\rm days}]$} &
\colhead{$\alpha_2$} &
\colhead{Ref.}}
\startdata
990123 & 1.600 & 1  &  $1.13\pm0.02$  &  $1.75\pm0.25$  &  $1.75\pm0.11$  & 2 \\
       & 1.600 & 3  &  $1.10\pm0.03$  &  $2.04\pm0.46$  &  $1.65\pm0.06$  & 4 \\
990510 & 1.619 & 5  &  $0.82\pm0.02$  &  $1.20\pm0.08$  &  $2.18\pm0.05$  & 6 \\
       &       &    &  $0.75\pm0.03$  &  $1.30\pm0.10$  &  $2.46\pm0.06$  & 7 \\
       &       &    &  $0.82\pm0.03$  &  $1.51\pm0.09$  &  $2.23\pm0.07$  & 8 \\
       &       &    &  $0.76\pm0.01$  &  $1.75\pm0.03$  &  $2.40\pm0.02$  & 9 \\
000301c& 2.03  &10  &  $0.92       $  &  $6.30       $  &  $3.11       $  &11 \\
       &       &    &  $0.72\pm0.34$  &  $4.39\pm1.52$  &  $2.29\pm1.00$  &12 \\  
\enddata
\tablerefs{
(1)  Andersen {\it et al.} 1999;
(2)  Castro-Tirado {\it et al.} 1999a;
(3)  Kelson {\it et al.} 1999; Hjorth {\it et al.} 1999;
(4)  Kulkarni {\it et al.} 1999; 
(5)  Vreeswijk {\it et al.} 1999b;
(6)  Harrison {\it et al.} 1999;               
(7)  Israel {\it et al.} 1999;                   
(8)  Beuermann {\it et al.} 1999;
(9)  Stanek {\it et al.} 1999;
(10) Castro {\it et al.} 2000;
(11) Rhoads \& Fruchter 2000;
(12) Jensen {\it et al.} 2000}
\tablecomments{$\alpha_1$ describes the early power-law decay, $\alpha_2$ 
the late-time power-law decay. $t_{\rm break}$ is the break time. }
\end{deluxetable}


\newpage 

\begin{references}

\reference{} Andersen, M. I., {\it et al.} 1999, Science, 283, 2075
\reference{} Aptekar, R., {\it et al.} 1995, Space Sci. Rev., 71, 265
\reference{} Beuermann, K., {\it et al.} 1999, A\&A, 352, L26
\reference{} Bloom, J. S., {\it et al.} 1999, Nature, 401, 453
\reference{} Bloom, J. S., {\it et al.} 2000, GCN Circ. 661 
\reference{} Calzetti, D., 1997, AJ, 113, 162
\reference{} Castro, S. M., {\it et al.} 2000, GCN Circ. 605 
\reference{} Castro-Tirado, A. J., \& Gorosabel, J. 1999, A\&AS 138, 449
\reference{} Castro-Tirado, A. J., {\it et al.} 1998, Science, 279, 1011
\reference{} Castro-Tirado, A. J., {\it et al.} 1999a, Science, 283, 2069
\reference{} Castro-Tirado, A. J., {\it et al.} 1999b, ApJ, 511, L85
\reference{} Castro-Tirado, A. J., {\it et al.} 2000, A\&A, submitted
\reference{} Chevalier, R. A., \& Li, Z.-Y. 1999, ApJ, 520, L29
\reference{} Chi, X., \& Wolfendale, A. W. 1991, J. Phys. G., 17, 987
\reference{} Dal Fiume, D., {\it et al.} 2000, A\&A, 355, 454
\reference{} Esin, A. A., \& Blandford, R. B. 2000, astro-ph/0003415
\reference{} Fishman, G. J., \& Meegan, C. A. 1995, 
             Annu. Rev. Astron. Astrophys., 33, 415
\reference{} Fixsen, D. J., Moseley, S. H., \& Arendt, R. G. 2000,  
             astro-ph/0002260
\reference{} Frail, D. A., 2000, GCN Circ. 655
\reference{} Frail, D. A., Waxman, E., \& Kulkarni, S. R. 2000, ApJ, 537, 191
\reference{} Fryer, C. L., Woosley, S. E., \& Hartmann, D. H. 
             1999, ApJ, 526, 152
\reference{} Galama, T. J., {\it et al.} 1998, ApJ, 500, L97
\reference{} Garcia, M. R., {\it et al.} 1998, ApJ, 500, L105
\reference{} Goldsten, J., {\it et al.} 1997, Space Sci. Rev., 82, 169
\reference{} Greiner, J. 2000, see http://www.aip.de/$^\sim$jcg/grbgen.html
\reference{} Hakkila, J., Myers, J., Stidham, B., \& Hartmann, D. H.
             1997, AJ, 114, 2043
\reference{} Halpern, J. P., Thorstensen, J. R., Helfand, D. J.,
             \& Costa, E. 1998, Nature, 393, 41
\reference{} Harrison, F. A., {\it et al.} 1999, ApJ, 523, L121
\reference{} Henden, A. A., 2000, GCN Circ. 662
\reference{} Henden, A. A., {\it et al.} 2000, GCN Circ. 652
\reference{} Hjorth, J., {\it et al.} 1999, GCN Circ. 219
\reference{} Hurley, K., {\it et al.} 1992, A\&AS, 92, 401
\reference{} Hurley, K., {\it et al.} 2000, ApJ, 534, L23
\reference{} Hurley, K., Cline, T., \& Mazets, E. 2000, GCN Circ. 642
\reference{} Israel, G. L., {\it et al.} 1999, A\&A, 348, L5
\reference{} Jensen, B. L., {\it et al.} 2000, astro-ph/0005609
\reference{} Kelson, D. D., {\it et al.} 1999, IAU Circ. 7096
\reference{} Klose, S. 2000, preprint astro-ph/0001008
\reference{} Klose, S., {\it et al.} 2000, GCN Circ. 645      
\reference{} Kulkarni, S. R., {\it et al.} 1998, Nature, 393, 35 
\reference{} Kulkarni, S. R., {\it et al.} 1999, Nature, 398, 389
\reference{} Lenzen, R., Bizenberger, P. Salm, N., \& Storz, C. 1998,
             SPIE, 3354, 493
\reference{} Masetti, N., {\it et al.} 2000, A\&A, 354, 473 
\reference{} M\'esz\'aros, P. 2000, Nucl. Phys. B, 80, 63 (Proceedings
             19th Texas Symposium, eds. \'E. Aubourg et al.)
\reference{} M\'esz\'aros, P., \& Rees, M. J., \& Wijers,  R. A. M. J. 1998,
             ApJ, 499, 301
\reference{} Metzger, M. R., \& Fruchter, A. 2000, GCN Circ. 669
\reference{} Mirabal, N. Halpern, J. P., Kemp, J., \& Helfand, D. J.
             2000, GCN Circ. 646
\reference{} Mirabal, N. Halpern, J. P., \& Wagner R. M.
             2000a, GCN Circ. 650
\reference{} Mirabal, N. Halpern, J. P., \& Wagner R. M.
             2000b, GCN Circ. 653
\reference{} Monet, D., {\it et al.} 1996, USNO-SA2.0, (U.S. Naval 
             Observatory, Washington DC)
\reference{} Paczy\'nski, B. 1998, ApJ, 494, L45
\reference{} Panaitescu, A., M\'esz\'aros, P., \& Rees, M. J. 1998, 
             ApJ, 503, 314
\reference{} van Paradijs, J., {\it et al.} 1997, Nature, 386, 686
\reference{} Palazzi, E., {\it et al.} 1998, A\&A, 336, L95
\reference{} Piran, T. 1999, Phys Rep., 314, 575
\reference{} Ramaprakash, A. N., {\it et al.} 1998, Nature, 393, 43
\reference{} Rhoads, J. E., \& Fruchter, A. S. 2000, astro-ph/0004057
\reference{} Rieke, G. H., \& Lebofsky, M. J. 1985, ApJ, 288, 618
\reference{} Sari, R., Piran, T., \& Halpern, R. 1999, ApJ, 519, L17
\reference{} Sari, R., Piran, T., \& Narayan, R. 1998, ApJ, 497, L17 
\reference{} Schlegel, D., Finkbeiner, D., \& Davis, M. 1998, ApJ, 500, 525
\reference{} Stanek, K. Z., {\it et al.} 1999, ApJ, 522, L39
\reference{} Stecklum, B., {\it et al.} 2000, GCN Circ. 654
\reference{} S\={u}d\v{z}ius, J., Bobinas, V., \& Raudeli\={u}nas, S. 1996,
             Baltic Astron., 5, 485
\reference{} Vrba, F. J., Coyne, G. V., \& Tapia, S. 1993, AJ, 105, 1010
\reference{} Vrba, F. J., {\it et al.} 2000, ApJ, 528, 254
\reference{} Vreeswijk, P. M., {\it et al.} 1999a, ApJ, 523, 171
\reference{} Vreeswijk, P. M., {\it et al.} 1999b, GCN Circ. 324
\reference{} Wainscoat, R. J., \& Cowie, L. L. 1992, AJ, 103, 332
\reference{} Wijers, R. A. M. J., Rees, M. J., \& M\'esz\'aros, P. 1997,
             MNRAS, 288, L51
\reference{} Zharikov, S. V., Sokolov, V. V., \& Baryshev, Yu. V. 1998,
             A\&A, 337, 356
\end{references}
\end{document}